\documentclass[twocolumn,showpacs,preprintnumbers,superscriptaddress,floatfix,nofootinbib]{revtex4}
\usepackage{amssymb}
\usepackage{amsmath}
\usepackage{graphicx}
\usepackage{dcolumn}
\usepackage{bm}
\usepackage[subfigure]{graphfig}
\usepackage{multirow}
\usepackage{makecell}
\usepackage{array}

\begin{document}

\title{Reexamination of the role of the $\Delta ^{\ast}$ resonances in the $pp\rightarrow
nK^{+}\Sigma ^{+}$ reaction}
\author{Xiao-Yun Wang\footnote{Electronic address: xywang@impcas.ac.cn}}
\affiliation{Institute of Modern Physics, Chinese Academy of Sciences, Lanzhou 730000,
China}
\affiliation{University of Chinese Academy of Sciences, Beijing 100049, China}
\affiliation{Research Center for Hadron and CSR Physics, Institute of Modern Physics of
CAS and Lanzhou University, Lanzhou 730000, China}
\author{Xu Cao\footnote{Correspondent: caoxu@impcas.ac.cn}}
\affiliation{Institute of Modern Physics, Chinese Academy of Sciences, Lanzhou 730000,
China}
\affiliation{Research Center for Hadron and CSR Physics, Institute of Modern Physics of
CAS and Lanzhou University, Lanzhou 730000, China}
\affiliation{State Key Laboratory of Theoretical Physics, Institute of Theoretical
Physics, Chinese Academy of Sciences, Beijing 100190, China}
\author{Ju-Jun Xie\footnote{Electronic address: xiejujun@impcas.ac.cn}}
\affiliation{Institute of Modern Physics, Chinese Academy of Sciences, Lanzhou 730000,
China}
\affiliation{Research Center for Hadron and CSR Physics, Institute of Modern Physics of
CAS and Lanzhou University, Lanzhou 730000, China}
\affiliation{State Key Laboratory of Theoretical Physics, Institute of Theoretical
Physics, Chinese Academy of Sciences, Beijing 100190, China}
\author{Xu-Rong Chen}
\affiliation{Institute of Modern Physics, Chinese Academy of Sciences, Lanzhou 730000,
China}
\affiliation{Research Center for Hadron and CSR Physics, Institute of Modern Physics of
CAS and Lanzhou University, Lanzhou 730000, China}

\begin{abstract}
In this work, the role of the $\Delta ^{\ast }$ resonances in the process of
$pp\rightarrow nK^{+}\Sigma ^{+}$ are systematically investigated with the
effective Lagrangian approach and the isobar model. We find that a $P_{31}$
state, either $\Delta ^{\ast }(1750)$ or $\Delta ^{\ast }(1910)$, is favored
by the data while the $P_{33}$ state, namely $\Delta ^{\ast }(1920)$, has
small contribution. Besides, either sub-threshold $S_{31}$ $\Delta ^{\ast
}(1620)$ resonance or strong $n\Sigma $ final state interaction or both have
possible contribution at near threshold region, depending on the measured cross sections.
We demonstrate the invariant
mass distributions and the Dalitz Plots in order to investigate whether it
is possible to distinguish the controversial $K\Sigma $ production mechanism
in these observables.
\end{abstract}

\maketitle

\section{Introduction}

\label{sec:intro}

The baryon spectrum has attracted a lot of theoretical and experimental
interest for a long time because it is expected to reveal important
information on the internal structure of baryons and the mechanism of quark
confinement. The phenomenological models \cite{sc86,ul01,ly96,ch02} predict
a rich of the excited states of $N^{\ast }$ and $\Delta ^{\ast }$, and recently
lattice QCD has been used to calculated the spectrum in finite volume\cite{jb10,sd08}.
However, although some of the
predicted states have been identified from the $\pi N$ and $\gamma N$
scattering data, many of them have not yet been observed in any experiments
\cite{Mandaglio:2010sg,Lleres:2008em,Gottschall:2013uha,Dugger:2013crn}. These states, so called as the \emph{missing resonances},
are what we are facing with and long seeking for \cite{rk80}. Therefore, it
is necessary and meaningful to search for these states and study their
properties in other reactions.

The $pp\rightarrow nK^{+}\Sigma^{+}$ reaction is a very ideal channel for
studying the $\Delta ^{\ast }$ resonances with isospin 3/2 since the
contributions of the $N^{\ast }$ with isospin 1/2 are filtered out in this channel.
Some results have been obtained on the experimental and theoretical aspects,
however, it is far from being sufficient to reveal the contribution of the $%
\Delta ^{\ast }$ on the basis of these results.

At present, there are only a few experimental data on the total cross
section of the $pp\rightarrow nK^{+}\Sigma ^{+}$ reaction \cite%
{ab88,tr06,yu07,yu10,ab10,yu11}. What is worse, it is known that the
close-to-threshold data are inconsistent between the COSY-11, HIRES and
COSY-ANKE groups. The total cross section data from COSY-11 shows strong
close-threshold enhancement \cite{tr06}, however, not confirmed by
the measurement of other two groups. The COSY-ANKE data follow the
behavior of three-body phase-space \cite{yu07,yu10} and the values are about
one order smaller than that of the COSY-11 at the same energy range \cite%
{tr06}. Moreover, the HIRES data \cite{ab10} at beam energy $T_{p}=2.08$ GeV
make the situation more complex and its value is around three times bigger
than the COSY-ANKE data at $T_{p}=2.16$ GeV \cite{yu07}. Valdau and Wilkin
argued that the HIRES data determined from the inclusive $K^{+}$-meson
production in $pp$ collisions should be considered as an upper bound so it
is not conflict with the result of COSY-ANKE \cite{yu11}.

On the theoretical side, most of the previous studies focus on the
contribution of the $\Delta ^{\ast }(1920)$ and $\Delta ^{\ast }(1620)$
resonances in the $pp\rightarrow nK^{+}\Sigma ^{+}$ reaction. Tsushima
\textit{et al.} introduced the effective intermediate $\Delta ^{\ast }(1920)$
resonance to account for the contribution of several $\Delta ^{\ast }$ state
around 1900 MeV \cite{kt94,kt97,kt99,as99,rs06} and their calculations
reproduced the experiment data at high energies very well. However, the
coupling of $\Delta ^{\ast }(1920)$ to the $K\Sigma$ in relative $P$%
-wave is suppressed at close-to-threshold energies. In order to explain the
large near-threshold data of COSY-11, Xie \textit{et al.} \cite{xie07}
suggested the $\Delta ^{\ast }(1620)$ resonance below the $K\Sigma$
threshold as the possible source of the very strong near-threshold
enhancement. Later, Cao \textit{et al.} \cite{cao08} further pointed out
that an unusual strong $n\Sigma$ final state interaction were needed to
fully interpret the COSY-11 data. In these calculations, the coupling
constant of the $\Delta ^{\ast }(1620)$ to $K\Sigma$ determined by the
relation $g_{\Delta ^{\ast }(1620)\Sigma K}=g_{\Delta ^{\ast }(1620)\pi N}$
from the SU(3) symmetry has big uncertainty because the mass of $\Delta
^{\ast }(1620)$ is below the $K\Sigma$ threshold.

The above situation indicates that the production mechanism of the $%
pp\rightarrow nK^{+}\Sigma ^{+}$ reaction is still an open question. As a
matter of fact, there is long discrepancy on various coupled-channel
study of the $\pi^+ p\rightarrow K^{+}\Sigma ^{+}$ reaction, where only the $%
\Delta ^{\ast }$ resonances are allowed, same as the $pp\rightarrow
nK^{+}\Sigma ^{+}$ channel. The Juelich model \cite{am00,am01,md11,Doring12}
finds that the $\Delta ^{\ast }(1620)$ is dominant in the low energies of
this reaction, while the Bonn-Gatchina partial wave analysis identifies the $%
\Delta ^{\ast }(1920)$ as the most essential contribution \cite{ava11,BoCh12}%
. The Giessen model with the K-matrix approximation claims the vital role of
the $\Delta ^{\ast }(1750)$ at close-threshold range \cite%
{gp02,Penner2,vs04,cao03}. The confusion is not relieved\cite%
{ava11,Penner2,cao03} in the
$\pi^- p\rightarrow K\Sigma$ and $\gamma N \to K\Sigma$ reactions
where the $N^*$ resonances are also contributing,
though more data are available there.
The situation at high energies is even more
complicated and several partial waves are important.

In this work, we systematically study the role of $\Delta ^{\ast++ }$
resonances in the $pp\rightarrow nK^{+}\Sigma ^{+}$ channel in order to properly
clarify the present confusion and shed light on the future measurements.
This paper is organized as follows. After the introduction, we illustrate
our investigative method and formalism. In Sec. III, the numerical results
are presented and discussed. We propose two possible schemes to interpret
the the contribution of $\Delta ^{\ast++ }$ resonances in the $pp\rightarrow
nK^{+}\Sigma ^{+}$ reaction. Finally, a short summary are given in Sec. IV.

\section{Method and formalism}

In the present work, we use the effective Lagrangian approach and the isobar
model in terms of hadrons to study the process of $pp\rightarrow
nK^{+}\Sigma ^{+}$ and $\pi ^{+}p\rightarrow K^{+}\Sigma ^{+}$, where the $%
K^{+}\Sigma ^{+}$ are produced through the intermediate $\Delta ^{\ast
}(1620)$, $\Delta ^{\ast }(1750)$, $\Delta ^{\ast }(1910)$ and $\Delta
^{\ast }(1920)$ resonances. Besides, the $\pi$-meson exchange in the $pp$
collisions are considered in the proton-proton collisions.
Other meson, e.g. the $\rho$-meson exchange, is not included and this is not unanimous in the modeling of the $pp \to nK^+\Sigma^+$ reaction within a meson-exchange picture. Fortunately, the estimation of the $pp \to nK^+\Sigma^+$ cross section in our model is sensitive to the couplings of different $\Delta ^{\ast }$ resonances to $K\Sigma$ channel, which are determined from the $\pi^+ p \to K^+\Sigma^+$ reaction. Hence, single-pion exchange is enough for this purpose. By neglecting the $\rho$-meson exchange, we can give a unified picture of pion- and proton-induced reactions, though our theoretical results are more general than this would suggest.

At present it is still under debate which $P_{31}$ state, the $\Delta
^{\ast}(1750)$ or $\Delta ^{\ast }(1910)$ resonance, have strong coupling to
$K\Sigma$, as discussed in Sec. \ref{sec:intro}. Based on the limited data
of the $pp\rightarrow nK^{+}\Sigma ^{+}$ reaction, it is impossible to
unambiguous pin down the relevant masses at this stage. So herein we
include these two $P_{31}$ states seperately, leading to two solutions
with different amplitudes,
\begin{eqnarray}
\mathcal{M}_{I}\text{ }\mathcal{=M}_{\Delta ^{\ast }(1620)}+\mathcal{M}%
_{\Delta ^{\ast }(1910)}+\mathcal{M}_{\Delta ^{\ast }(1920)}
\label{eq:solution1}
\\
\mathcal{M}_{II}\text{ }\mathcal{=M}_{\Delta ^{\ast }(1620)}+\mathcal{M}%
_{\Delta ^{\ast }(1750)}+\mathcal{M}%
_{\Delta ^{\ast }(1920)}
\label{eq:solution2}
\end{eqnarray}%
as summarized in Tab.~\ref{tab:sumr}. This is also in line with the study of the $\pi^+ p\rightarrow K^{+}\Sigma
^{+}$ reaction in different models \cite%
{am00,am01,md11,Doring12,ava11,BoCh12,gp02,Penner2,vs04,cao03}, which
usually include only one of the $P_{31}$ states.
Correspondingly we will consider these two solutions in the $\pi^+ p\rightarrow K^{+}\Sigma
^{+}$ reaction in the following calculation.

\begin{table}[t]
\begin{minipage}{0.5\textwidth}\centering
\caption{The considered $\Delta ^{\ast }$ resonances in the model} \label{tab:sumr}%
\begin{tabular}{|c|c|c|c|c|}
\Xhline{1.2pt}\hline\centering
  Resonances            & Width (MeV) & $J^P$   & Solution I & Solution II \\
  \Xhline{0.8pt}
$\Delta ^{\ast }(1620)S_{31}$ & 140         & $1/2^-$ & \checkmark & \checkmark \\
$\Delta ^{\ast }(1750)P_{31}$ & 300         & $1/2^+$ & ---        & \checkmark  \\
$\Delta ^{\ast }(1910)P_{31}$ & 250         & $1/2^+$ & \checkmark & ---        \\
$\Delta ^{\ast }(1920)P_{33}$ & 220         & $3/2^+$ & \checkmark & \checkmark  \\
  \Xhline{0.8pt}
\end{tabular}
\end{minipage}
\end{table}

\subsection{Feynman diagrams and effective Lagrangian}

\begin{figure}[ht]
\centering
\includegraphics[height=5.4cm,width=0.45\textwidth]{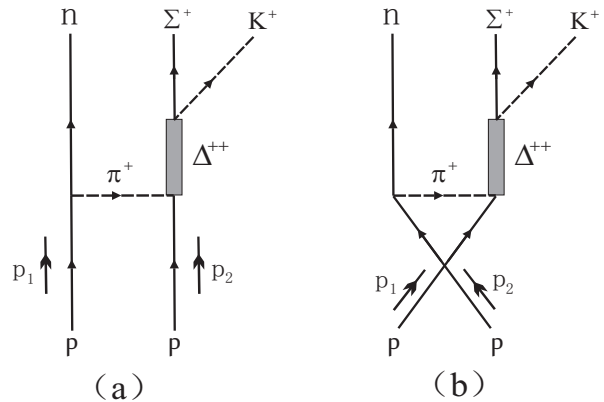}
\caption{Feynman diagram for the $pp\rightarrow nK^{+}\Sigma ^{+}$ reaction.}
\label{ppfeymn}
\end{figure}

\begin{figure}[ht]
\centering
\includegraphics[height=4.0cm,width=0.45\textwidth]{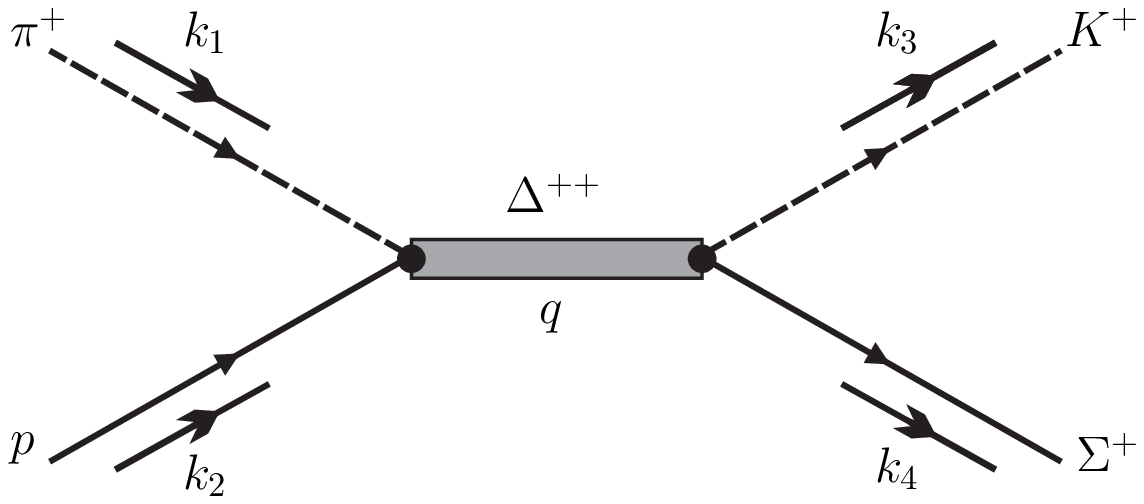}
\caption{Feynman diagram for the $\protect\pi ^{+}p\rightarrow K^{+}\Sigma
^{+}$ reaction.}
\label{pipfeymn}
\end{figure}

The basic tree-level Feynman diagrams for the $pp\rightarrow nK^{+}\Sigma
^{+}$ reaction are presented in Fig. \ref{ppfeymn}, and the $s$-channel
diagram for the $\pi ^{+}p\rightarrow K^{+}\Sigma ^{+}$ reaction is
depicted in Fig. \ref{pipfeymn}. The $t$-channel diagram for the $\pi
^{+}p\rightarrow K^{+}\Sigma ^{+}$ reaction is calculated to be small \cite%
{cao03}. This is reasonable because the exchanged $K$ and $K^*$ mesons in the
$t$-channel has small coupling to the relevant $N\Sigma$ and $\pi K$ channels.
The interference of $u$- and $t$-channel with $s$-channel resonances contribution
are important for describing
the differential and polarization observables \cite{cao03},
but it is safe to ignore them in the determination of the coupling constants
of the dominant resonances in $s$-channel.

For the interaction vertex of $\pi NN$, we use the effective pseudoscalar
coupling \cite{kt94,kt97,kt99}

\begin{equation}
\mathcal{L}_{\pi NN}=-ig_{\pi NN}\bar{N}\gamma _{5}\vec{\tau}\cdot \vec{\pi}N
\end{equation}

The Lagrangians of $\Delta ^{\ast }N\pi $ and $\Delta ^{\ast
}K\Sigma $ vertices are used by many models, such as J\"{u}lich model,
Giessen model and Bonn-Gatchina model \cite{md11,gp02,ava11}. But the
elementary Lorentz structure which depends on the relative orbital momentum
and spin are the same. Therefore, the general effective Lagrangian
for the vertices of $\Delta ^{\ast }N\pi $ and $\Delta ^{\ast }\Sigma K $ read
as follows:

\begin{equation}
\mathcal{L}_{_{\Delta ^{\ast }(1620)N\pi }}=\frac{g_{\Delta ^{\ast
}(1620)N\pi }}{m_{\pi }}\bar{\Delta}^{\ast }\gamma ^{\mu }\vec{\tau}\cdot
\partial _{\mu }\vec{\pi}N+h.c.
\end{equation}

\begin{equation}
\mathcal{L}_{_{\Delta ^{\ast }(1620)\Sigma K}}=\frac{g_{\Delta ^{\ast
}(1620)\Sigma K}}{m_{K}}\bar{\Delta}^{\ast }\gamma ^{\mu }\vec{\tau}\cdot
\partial _{\mu }\vec{K}\Sigma +h.c.
\end{equation}

\begin{equation}
\mathcal{L}_{_{\Delta ^{\ast }(1750)N\pi }}=-\frac{g_{\Delta ^{\ast
}(1750)N\pi }}{m_{\Delta ^{\ast }(1750)}}\bar{\Delta}^{\ast }\gamma
_{5}\gamma _{\mu }\vec{\tau}\cdot \partial ^{\mu }\vec{\pi}N+h.c.
\end{equation}

\begin{equation}
\mathcal{L}_{_{\Delta ^{\ast }(1750)\Sigma K}}=-\frac{g_{\Delta ^{\ast
}(1750)\Sigma K}}{m_{\Delta ^{\ast }(1750)}}\bar{\Delta}^{\ast }\gamma
_{5}\gamma _{\mu }\vec{\tau}\cdot \partial ^{\mu }\vec{K}\Sigma +h.c.
\end{equation}%

\begin{equation}
\mathcal{L}_{_{\Delta ^{\ast }(1910)N\pi }}=-\frac{g_{\Delta ^{\ast
}(1910)N\pi }}{m_{\Delta ^{\ast }(1910)}}\bar{\Delta}^{\ast }\gamma
_{5}\gamma _{\mu }\vec{\tau}\cdot \partial ^{\mu }\vec{\pi}N+h.c.
\end{equation}

\begin{equation}
\mathcal{L}_{_{\Delta ^{\ast }(1910)\Sigma K}}=-\frac{g_{\Delta ^{\ast
}(1910)\Sigma K}}{m_{\Delta ^{\ast }(1910)}}\bar{\Delta}^{\ast
}\gamma _{5}\gamma _{\mu }\vec{\tau}\cdot \partial ^{\mu }\vec{K}\Sigma
+h.c.
\end{equation}

\begin{equation}
\mathcal{L}_{_{\Delta ^{\ast }(1920)N\pi }} =-\frac{g_{\Delta ^{\ast
}(1920)N\pi }}{m_{\Delta ^{\ast }(1920)}}\bar{\Delta}_{\mu }^{\ast }\vec{\tau%
}\cdot \partial ^{\mu }\vec{\pi}N+h.c.
\end{equation}

\begin{equation}
\mathcal{L}_{_{\Delta ^{\ast }(1920)\Sigma K}}=-\frac{g_{\Delta ^{\ast
}(1920)\Sigma K}}{m_{\Delta ^{\ast }(1920)}}\bar{\Delta}_{\mu }^{\ast }\vec{%
\tau}\cdot \partial ^{\mu }\vec{K}\Sigma +h.c
\end{equation}

where\ $\vec{\tau}$ is the Pauli matrix, and $\Delta ^{\ast }$ and $\Delta
_{\mu }^{\ast }$ stand for the fields of the corresponding baryon resonances.

\subsection{Propagator and Form factor}

The propagator of the $\pi $-meson is,%
\begin{equation}
G_{\pi }(q_{\pi })=\frac{-i}{q_{\pi }^{2}-m_{\pi }^{2}}\quad .
\end{equation}

The propagators for the resonance $\Delta ^{\ast }$ can be constructed
through projection operator and Breit-Wigner factor \cite{wh02}. For the $%
\Delta ^{\ast }(1620)$, $\Delta ^{\ast }(1750)$ and $\Delta ^{\ast }(1910)$
with spin-1/2, the propagator can be written as,

\begin{equation}
G_{\Delta ^{\ast }}^{\frac{1}{2}}(q_{\Delta ^{\ast }})=-i\frac{%
\rlap{$\slash$}q_{\Delta ^{\ast }}+m_{\Delta ^{\ast }}}{q_{\Delta ^{\ast
}}^{2}-m_{\Delta ^{\ast }}^{2}+im_{\Delta ^{\ast }}\Gamma _{\Delta ^{\ast }}}%
\quad .
\end{equation}

For $\Delta ^{\ast }(1920)$ with spin-3/2, we have
\begin{eqnarray}
G^{3/2}_{\Delta ^{\ast }}(q_{\Delta ^{\ast }})=G^{1/2}_{\Delta ^{\ast }}(q_{\Delta ^{\ast }}) G_{\mu\nu}(q_{\Delta ^{\ast }})
\\
G_{\Delta ^{\ast }}^{\mu \nu }(q_{\Delta ^{\ast }}) =-i\frac{\rlap{$\slash$}%
q_{\Delta ^{\ast }}+m_{\Delta ^{\ast }}}{q_{\Delta ^{\ast }}^{2}-m_{\Delta
^{\ast }}^{2}+im_{\Delta ^{\ast }}\Gamma _{\Delta ^{\ast }}}  \notag \\
\times \left[ g^{\mu \nu }-\frac{1}{3}\gamma ^{\mu }\gamma ^{\nu }-\frac{%
(\gamma ^{\mu }q_{\Delta ^{\ast }}^{\nu }-\gamma ^{\nu }q_{\Delta ^{\ast
}}^{\mu })}{3m_{\Delta ^{\ast }}}-\frac{2q_{\Delta ^{\ast }}^{\mu }q_{\Delta
^{\ast }}^{\nu }}{3m_{\Delta ^{\ast }}^{2}}\right]
\end{eqnarray}

At each vertex a relevant off-shell form factor is used to suppress the
contributions from high exchanged momenta. In our computation, we take the same form
factors as that used in the well-known Bonn model for the $\pi NN$ and $%
\Delta ^{\ast }N\pi $ vertices \cite{rb90}

\begin{eqnarray}
F_{\pi }^{NN}(q_{\pi }^{2})=\frac{\Lambda _{\pi }^{2}-m_{\pi }^{2}}{\Lambda
_{\pi }^{2}-q_{\pi }^{2}}\quad ,\\
F_{\pi }^{\Delta ^{\ast }N}(q_{\pi }^{2})=\frac{\Lambda _{\pi }^{\ast
2}-m_{\pi }^{2}}{\Lambda _{\pi }^{\ast 2}-q_{\pi }^{2}}\quad ,
\end{eqnarray}
where $q_{\pi }$ and $\Lambda _{\pi }^{(\ast)}$ are the four-momentum, and
cut-off parameters for the exchange $\pi $-meson, respectively. We take $\Lambda
_{\pi } =$ 0.8 GeV for all resonances and $\Lambda _{\pi }^{\ast }$ = 0.8 GeV, 1.0 GeV, 1.2 GeV
and 1.2 GeV for the $\Delta ^{\ast }(1620)$, $\Delta ^{\ast }(1750)$, $%
\Delta ^{\ast }(1910)$ and $\Delta ^{\ast }(1920)$ resonances, respectively.
They are determined by the data of $pp\rightarrow
nK^{+}\Sigma ^{+}$. The $\Lambda _{\pi }^{\ast }$ of $\Delta ^{\ast }(1620)$  can be determined
in the close-to-threshold region while those of the $\Delta ^{\ast }(1750)$ and $%
\Delta ^{\ast }(1910)$ can be pinned down at higher energies.
The uncertainty of the $\Lambda _{\pi }^{\ast }$ for $\Delta ^{\ast }(1920)$ is relatively bigger
because its contribution is small. For consistency, we set it to be the same as that of $%
\Delta ^{\ast }(1910)$.

Besides, the form factor for the off-shell resonances is taken as follows,%
\begin{equation}
F_{\Delta ^{\ast }}(q_{\Delta ^{\ast }}^{2})=\frac{\Lambda _{\Delta ^{\ast
}}^{4}}{\Lambda _{\Delta ^{\ast }}^{4}+(q^{2}-m_{\Delta ^{\ast }}^{2})^{2}} \quad ,
\end{equation}%
which is used to depict the resonances in the $\pi ^{+}p\rightarrow K^{+}\Sigma ^{+}$ and $%
pp\rightarrow nK^{+}\Sigma^{+} $ reactions. The revelent cut-off parameters
$\Lambda _{\Delta ^{\ast }}=1.7$ GeV are taken to be around the mass of resonances in both reactions
and the calculated results are not very sensitive to this value.

\subsection{Coupling constants}

The coupling constant of the $\pi NN$ interaction was given in many
theoretical works, and we take $g_{\pi NN}^{2}/4\pi =12.96$ \cite{lin00,Baru:2011NPA}.
According to above Lagrangians, the partial decay widths which are related
to the coupling constants can be written as follows:

\begin{eqnarray}
\Gamma _{\Delta ^{\ast }(1620)\rightarrow N\pi } &=&\frac{g_{\Delta ^{\ast
}(1620)N\pi \notag}^{2}(E_{N}+m_{N})\left\vert \vec{p}_{N}^{~\mathrm{c.m.}%
}\right\vert }{4\pi m_{\Delta ^{\ast }(1620)}m_{\pi }^{2}} \notag \\
&&\times (m_{\Delta ^{\ast }(1620)}-m_{N})^{2}\quad ,
\\
\Gamma _{\Delta ^{\ast }(1750)\rightarrow N\pi } &=&\frac{g_{\Delta ^{\ast
}(1750)N\pi }^{2}(E_{N}-m_{N})\left\vert \vec{p}_{N}^{~\mathrm{c.m.}%
}\right\vert }{4\pi m_{\Delta ^{\ast }(1750)}^{3}}  \notag \\
&&\times (m_{\Delta ^{\ast }(1750)}+m_{N})^{2}\quad ,
\\
\Gamma _{\Delta ^{\ast }(1910)\rightarrow N\pi } &=&\frac{g_{\Delta ^{\ast
}(1910)N\pi }^{2}(E_{N}-m_{N})\left\vert \vec{p}_{N}^{~\mathrm{c.m.}%
}\right\vert }{4\pi m_{\Delta ^{\ast }(1910)}^{3}}  \notag \\
&&\times (m_{\Delta ^{\ast }(1910)}+m_{N})^{2}\quad ,
\\
\Gamma _{\Delta ^{\ast }(1920)\rightarrow N\pi } &=& \frac{g_{\Delta ^{\ast
}(1920)N\pi }^{2}(E_{N}+m_{N})}{12\pi m_{\Delta ^{\ast }(1920)}^{3}}%
\left\vert \vec{p}_{N}^{~\mathrm{c.m.}}\right\vert ^{3}
\end{eqnarray}%
where the $E_{N}$, $E_{\pi }$ and $\vec{p}_{N}^{~\mathrm{c.m.}}$ are defined
in the center of mass (c.m.) system:

\begin{eqnarray*}
E_{N} &=& \frac{M_{\Delta^{\ast }}^{2}+ m_{N}^{2} - m_{\pi }^{2}}{2M_{\Delta ^{\ast }}}\quad , \\
\left\vert \vec{p}_{N}^{~\mathrm{c.m.}}\right\vert &=& \sqrt{E_{N}^{2} - m_{N}^{2} } \quad .
\end{eqnarray*}

For the $\Delta ^{\ast }\rightarrow K\Sigma$ decays, the formulae are basically
identical as those for the $\Delta ^{\ast }\rightarrow \pi N$ with the
replacement of $\pi$ and $N$ to $K$ and $\Sigma$, respectively. With the experimental
masses, total decay widths and branching ratios \cite{pdg}, we can obtain
all relevant $\Delta ^{\ast }$ resonance parameters from above formulae as
summarized in Table II.
\begin{table*}[t]
\begin{minipage}{1\textwidth}\centering
\caption{Relevant parameters for $\Delta ^{\ast }$ resonances. The values labeled as $^\dag$ are
extracted from the data of $\pi^{+}p\rightarrow K^{+}\Sigma^{+}$ reaction and others are from the compilation of PDG \cite{pdg}.}
\begin{tabular}{|p{2cm}<{\centering}|p{2cm}<{\centering}|p{2cm}<{\centering}|p{2cm}<{\centering}
|p{3cm}<{\centering}|p{2cm}<{\centering}|p{2cm}<{\centering}|}
\Xhline{1.2pt}\hline\centering
Resonances & mass (MeV) & width(MeV) & channel & Branching ratio (\%) & $g^{2}/4\pi
$ \\ \Xhline{0.8pt}
$\Delta ^{\ast }(1620)$ & 1615 & 140 & $\pi N$ & 25.0 & 0.002 \\ \cline{4-6}
&  & & $K\Sigma $ & - & 0.053$^\dag$ \\ \Xhline{0.8pt}
$\Delta ^{\ast }(1750)$ & 1750 & 300 & $\pi N$ & 10.0 & 0.20 \\ \cline{4-6}
&  & & $K\Sigma $ & 7.1 &2.96$^\dag$ \\  \hline \Xhline{1.2pt}
$\Delta ^{\ast }(1910)$ & 1875 & 250 & $\pi N$ & 22.5 & 0.288 \\ \cline{4-6}
&  & & $K\Sigma $ & 14.0 & 0.953 \\ \Xhline{0.8pt}
$\Delta ^{\ast }(1920)$ & 1910 & 220 & $\pi N$ & 12.5 & 0.730 \\ \cline{4-6}
&  & & $K\Sigma $ & 2.14 & 0.510 \\ \Xhline{0.8pt}
\end{tabular}
\end{minipage}
\end{table*}
In this table, all the known branching ratios of the $\Delta ^{\ast }(1620)$,
$\Delta ^{\ast }(1750)$, $\Delta ^{\ast }(1910)$ and $%
\Delta ^{\ast }(1920)$ resonances are taken from
the Particle Data Group (PDG) \cite{pdg}.

Since the mass of the $\Delta
^{\ast }(1620)$ is below the threshold of the $K\Sigma ,$ the coupling of the $%
\Delta ^{\ast }(1620)$ to $K\Sigma $ can not be determined by the corresponding decay width.
Also, there is no so much information on the coupling strength of the $\Delta
^{\ast }(1750)K\Sigma $ vertex. In our calculation, they are treated as free parameters
and fitted to the data of $\protect\pi ^{+}p\rightarrow K^{+}\Sigma
^{+}$ reaction. Following the Feynman rules
and using above Lagrangian, the theoretical invariant amplitude $\mathcal{A%
}$ of $\pi ^{+}p\rightarrow K^{+}\Sigma ^{+}$ reaction
in Fig. 2 could be calculated as,%
\begin{eqnarray}\label{eq:amppiN}
\mathcal{A}_{I} &\mathcal{=}& \frac{g_{\Delta ^{\ast }(1620)N\pi }g_{\Delta
^{\ast }(1620)\Sigma K}F_{\Delta ^{\ast }(1620)}(q_{\Delta ^{\ast }}^{2})}{%
m_{K}m_{\pi}} \notag\\
&&\times \bar{u}(k_{4})\rlap{$\slash$}k_{3}G_{\Delta ^{\ast
}(1750)}(q_{\Delta ^{\ast }}^{2})\rlap{$\slash$}k_{1}u(k_{2})\notag\\
&+& \frac{g_{\Delta ^{\ast }(1750)N\pi }g_{\Delta
^{\ast }(1750)\Sigma K}F_{\Delta ^{\ast }(1750)}(q_{\Delta ^{\ast }}^{2})}{%
m_{\Delta ^{\ast }(1750)}^{2}} \notag\\
&&\times \bar{u}(k_{4})\gamma _{5}\rlap{$\slash$}k_{3}G_{\Delta ^{\ast
}(1750)}(q_{\Delta ^{\ast }}^{2})\rlap{$\slash$}k_{1}\gamma _{5}u(k_{2})\notag\\
&+& \frac{g_{\Delta ^{\ast }(1920)N\pi }g_{\Delta
^{\ast }(1920)\Sigma K}F_{\Delta ^{\ast }(1920)}(q_{\Delta ^{\ast }}^{2})}{%
m_{\Delta ^{\ast }(1920)}^{2}} \notag\\
&&\times \bar{u}(k_{4}) k_{3\mu}G^{\mu \nu }_{\Delta ^{\ast
}(1920)}(q_{\Delta ^{\ast }}^{2})k_{1\nu}u(k_{2})\quad ,
\end{eqnarray}%
if assuming the intermediate $P_{31}$ excitation is the $\Delta ^{\ast }(1750)$ resonance.
Here the propagator $G_{\Delta ^{\ast }}$ and the form factor $%
F_{\Delta^{\ast }}$ of the $\Delta ^{\ast }$ resonance can be
found in the following subsection. By integrating the amplitude
in the two-body phase space, we can easily obtain
the total cross sections of the $\pi ^{+}p\rightarrow K^{+}\Sigma ^{+}$
reaction as function of the momentum of beam particle $\pi ^{+}$-meson.
By fitting the coupling constants od $\Delta ^{\ast
}(1620)K\Sigma $ and $\Delta ^{\ast
}(1750)K\Sigma $, we achieve a good agreement ($\chi^2 = 3.6$) between the
model and the experimental data, as shown in Fig. 3(a) and Tab. II. The extracted
parameter $g_{\Delta
^{\ast }(1750)\Sigma K}^{2}/4\pi =2.96$ gives a reasonable branch
ratio 7.1\% of $\Delta ^{\ast }(1750)\rightarrow K\Sigma $, which is around
one order larger than that in the refined Giessen model (0.9\%) \cite{cao03}%
. However, it should be noted the mass and total width of $\Delta ^{\ast
}(1750)$ are different in two approaches. Our $g_{\Delta
^{\ast }(1620)\Sigma K}^{2}/4\pi =0.053$
is about one order smaller than the value from SU(3) symmetry in Ref. \cite{xie07},
but in the same level with the value of Giessen model \cite{cao03}.
In an alternative explanation of the $\pi ^{+}p\rightarrow K^{+}\Sigma ^{+}$ data,
the $\Delta ^{\ast }(1750)$ would be replaced by the $\Delta ^{\ast }(1910)$
in Eq. (\ref{eq:amppiN}), corresponding to the amplitudes $\mathcal{A}_{II}$ in solution II ($\chi^2 = 4.6$).
The calculated total cross section of $\pi ^{+}p\rightarrow K^{+}\Sigma ^{+}$
with the parameters in Tab. II are shown in Fig. 3(b). As can be seen, the
two solutions both give a fair reproduction of the data, reflecting the
validity and consistency of our parameters.

\begin{figure}[t]
\centering
\includegraphics[height=5.5cm,width=0.48\textwidth]{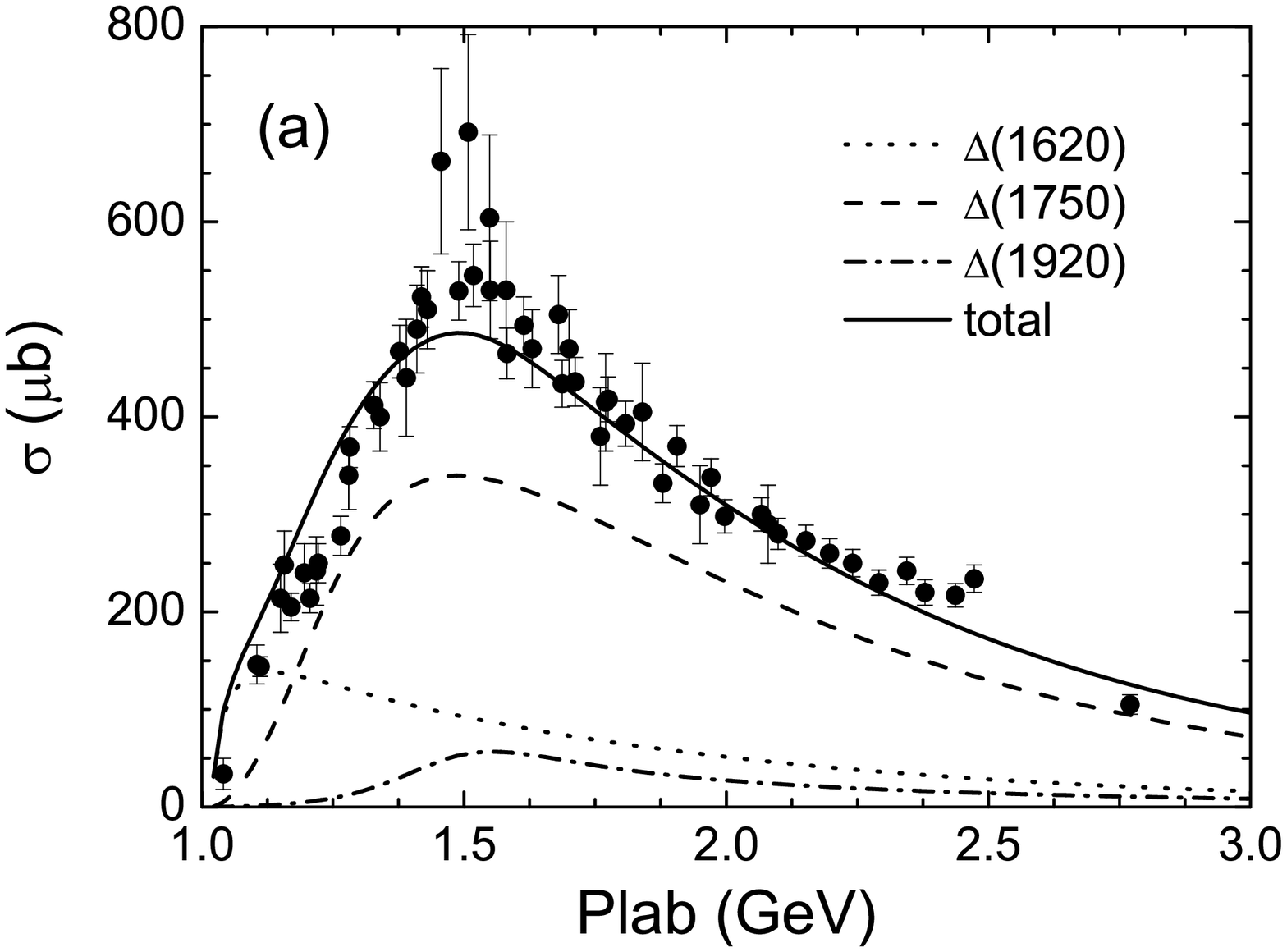}
\includegraphics[height=5.5cm,width=0.48\textwidth]{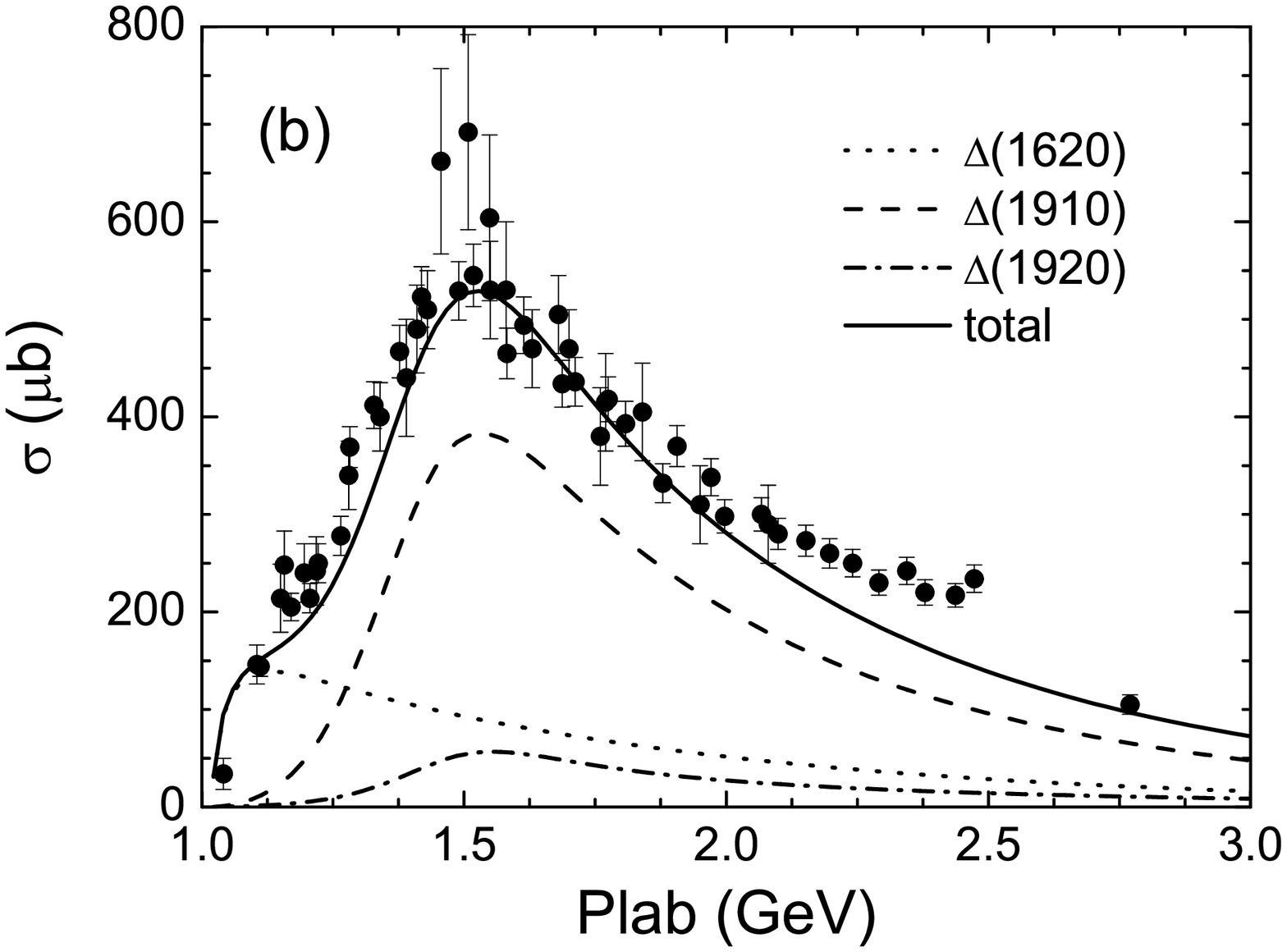}
\caption{Total cross section including the contributions of $\Delta ^{\ast
}(1750)$ resonance versus the beam momentum P$_{lab}$ for $\protect\pi %
^{+}p\rightarrow K^{+}\Sigma ^{+}$ reaction. The experimental data are taken
from Ref. \protect\cite{ab88}.}
\end{figure}

\subsection{Amplitude}

According to above effective Lagrangian and the Feynman rules, the invariant
amplitudes of the $\Delta ^{\ast }(1620)$, $\Delta ^{\ast }(1750)$, $\Delta
^{\ast }(1910)$ and $\Delta ^{\ast }(1920)$ resonances contribution in the $%
pp\rightarrow nK^{+}\Sigma ^{+}$ reaction could be read as,

\begin{eqnarray}
\mathcal{M}_{\Delta ^{\ast }(1620)} &=&\frac{\sqrt{2}g_{\pi NN}g_{\Delta
^{\ast }(1620)\Sigma K}g_{\Delta ^{\ast }(1620)N\pi }}{m_{K}m_{\pi }}  \notag
\\
&&\times F_{\pi }^{NN}(q_{\pi }^{2})F_{\pi }^{\Delta ^{\ast }N}(q_{\pi
}^{2})F_{\Delta ^{\ast }}(q_{\Delta ^{\ast }}^{2})  \notag \\
&&\times \bar{u}_{\Sigma }(p_{\Sigma })\rlap{$\slash$}p_{K}G_{\Delta ^{\ast
}}^{\frac{1}{2}}(q_{\Delta ^{\ast }})\rlap{$\slash$}p_{\pi }  \notag \\
&&\times u_{N}(p_{1})G_{\pi }(q_{\pi })\bar{u}_{N}(p_{n})\gamma
_{5}u_{N}(p_{2})
\\
\mathcal{M}_{\Delta ^{\ast }(1750)} &=&\frac{\sqrt{2}g_{\pi NN}g_{\Delta
^{\ast }(1750)N\pi }g_{\Delta ^{\ast }(1750)\Sigma K}}{m_{\Delta ^{\ast
}(1750)}^{2}}  \notag \\
&&\times F_{\pi }^{NN}(q_{\pi }^{2})F_{\pi }^{\Delta ^{\ast }N}(q_{\pi
}^{2})F_{\Delta ^{\ast }}(q_{\Delta ^{\ast }}^{2})  \notag \\
&&\times \bar{u}_{\Sigma }(p_{\Sigma })\gamma _{5}\rlap{$\slash$}%
p_{K}G_{\Delta ^{\ast }}^{\frac{1}{2}}(q_{\Delta ^{\ast }})%
\rlap{$\slash$}p_{\pi } \gamma _{5} \notag \\
&&\times u_{N}(p_{1})G_{\pi }(q_{\pi })\bar{u}_{N}(p_{n})\gamma
_{5}u_{N}(p_{2})
\\
\mathcal{M}_{\Delta ^{\ast }(1910)} &=&\frac{\sqrt{2}g_{\pi NN}g_{\Delta
^{\ast }(1910)N\pi }g_{\Delta ^{\ast }(1910)\Sigma K}}{m^2_{\Delta ^{\ast
}}(1910)}  \notag \\
&&\times F_{\pi }^{NN}(q_{\pi }^{2})F_{\pi }^{\Delta ^{\ast }N}(q_{\pi
}^{2})F_{\Delta ^{\ast }}(q_{\Delta ^{\ast }}^{2})  \notag \\
&&\times \bar{u}_{\Sigma }(p_{\Sigma })\gamma _{5}\rlap{$\slash$}%
p_{K}G_{\Delta ^{\ast }}^{\frac{1}{2}}(q_{\Delta ^{\ast }})%
\rlap{$\slash$}p_{\pi } \gamma _{5} \notag \\
&&\times u_{N}(p_{1})G_{\pi }(q_{\pi })\bar{u}_{N}(p_{n})\gamma
_{5}u_{N}(p_{2})
\\
\mathcal{M}_{\Delta ^{\ast }(1920)} &=&\frac{\sqrt{2}g_{\pi NN}g_{\Delta
^{\ast }(1910)N\pi }g_{\Delta ^{\ast }(1910)\Sigma K}}{m_{\Delta ^{\ast
}(1920)}^{2}}  \notag \\
&&\times F_{\pi }^{NN}(q_{\pi }^{2})F_{\pi }^{\Delta ^{\ast }N}(q_{\pi
}^{2})F_{\Delta ^{\ast }}(q_{\Delta ^{\ast }}^{2})  \notag \\
&&\times \bar{u}_{\Sigma }(p_{\Sigma })(p_{K})_{\mu }G_{\Delta ^{\ast
}}^{\mu \nu }(q_{\Delta ^{\ast }})(p_{\pi })_{\nu }  \notag \\
&&\times u_{N}(p_{1})G_{\pi }(q_{\pi })\bar{u}_{N}(p_{n})\gamma
_{5}u_{N}(p_{2})
\end{eqnarray}%
where $u_{\Sigma }$ and $u_{N}$ are the dirac wave functions of the $\Sigma $
baryon and the nucleon, respectively. The $p_{1}$ and $p_{2}$ denote the 4-momentum of the
initial protons. The above amplitudes are for the diagrams depicted in Fig.
1(a). For the Fig. 1(b), we only need to exchange $%
p_{1}$ with $p_{2}$ in the above formula.

The influence of the $n\Sigma^+$ final state interaction
(FSI) on the near- threshold behaviour is possibly weaker than $%
N\Lambda$ interaction as suggested in the literature \cite{yu07,yu10,yu11}.
This FSI effect, instead of the
sub-threshold $\Delta ^{\ast }(1620)$, would give the near threshold
enhancement in the total cross section. This gives rise to alternative
solutions of solution I and II. For the moment we do not have detailed
information on this $n\Sigma^+$ FSI, so we do not know the magnitude of the
impact of this FSI on the total cross section. For these reasons we simply
factor the amplitudes as \cite{ml76},
\begin{eqnarray}
\mathcal{M^{\prime }}_{I}\text{ }\mathcal{=(M}_{\Delta ^{\ast }(1620)} +%
\mathcal{M}_{\Delta ^{\ast }(1910)}+
\mathcal{M}_{\Delta ^{\ast }(1920)})T_{n\Sigma } \quad  \label{eq:solution1FSI}
\\
\mathcal{M^{\prime }}_{II}\text{ }\mathcal{=(M}_{\Delta ^{\ast }(1620)} +%
\mathcal{M}_{\Delta ^{\ast }(1750)}+\mathcal{M}_{\Delta ^{\ast }(1920)})
T_{n\Sigma } \quad \label{eq:solution2FSI}
\end{eqnarray}
The $T_{n\Sigma }$ is the Jost function describing the $n\Sigma ^{+}$ final
state interaction and goes to unity if no FSI. Analogy to the $p\Lambda $
FSI in $pp\rightarrow pK^{+}\Lambda $ reaction \cite{as06}, we take the same
formular to depict the $T_{n\Sigma }$ as used in Ref. \cite{xie07}:%
\begin{equation*}
T_{n\Sigma }=\frac{q+i\beta }{q-i\alpha }
\end{equation*}%
where $q$ is the internal momentum of $n$-$\Sigma ^{+}$ subsystem. Adjusting
our numerical calculations to the experiment data and also referring the $%
p\Lambda $ interaction in $pp\rightarrow pK^{+}\Lambda $ reaction \cite{as06}%
, the values of the $\alpha $ and $\beta $ are chose to be,%
\begin{equation*}
\alpha =-70\text{ MeV\quad , \ }\beta =280\text{ MeV\quad .}
\end{equation*}%
The  scattering length and effective range can be calculated by $\alpha $ and $\beta ,$%
\begin{equation*}
a=\frac{\alpha +\beta }{\alpha \beta }\text{\quad , \ \ \ }r=\frac{2}{\alpha
+\beta }\quad .
\end{equation*}%
The above values of $\alpha $ and $\beta $ correspond to the scattering
length $a=2.1$ fm and effective range $r=1.9$ fm, which is close to the $%
a=1.6$ fm and $r=3.2$ fm in Ref. \cite{xie07}.

In our model, the initial state interaction (ISI) is not considered because
it is difficult to treat the ISI unambiguously due to the lack of the
accurate NN interaction model at such high incident beam energies.
Hanhart and Nakayama \cite{Hanhart:1998rn} claims that the ISI has practically little influence on the energy dependence of the meson production cross section of nucleon- nucleon collisions close to threshold, and the reduction factor to the cross section can be roughly estimated by the $NN$ phase shifts and inelasticities. In our paper, we do not consider this reduction factor because this estimation is rough so it would cause uncertainty in the model. In fact, the cut-off values in form factors partly play the role of this reduction factor, as prescribed in previous studies of nucleon-nucleon collisions \cite{kt94,kt97,kt99,as99,rs06,xie07,cao08,am00,am01}. This is possibly the reason that the used cut-off values are smaller than the usual ones.

The total cross section of the $pp\rightarrow nK^{+}\Sigma ^{+}$ reaction
could be integrate the invariant amplitudes in the three-body phase space,

\begin{eqnarray}
d\sigma (pp &\rightarrow &nK^{+}\Sigma ^{+})=\frac{m_{p}^{2}}{\sqrt{%
(p_{1}\cdot p_{2})-m_{p}^{4}}}\left( \frac{1}{4}\sum\limits_{spins}\left%
\vert \mathcal{M}\right\vert ^{2}\right)  \notag \\
&&\times (2\pi )^{4}d\Phi _{3}(p_{1}+p_{2};p_{n},p_{K},p_{\Sigma })
\end{eqnarray}%
where the three-body phase space is defined as \cite{pdg}%
\begin{equation}
d\Phi _{3}=4 m_n m_{\Sigma}\delta ^{4}\left(
p_{1}+p_{2}-\sum\limits_{i=1}^{3}p_{i}\right) \prod\limits_{i=1}^{3}\frac{%
d^{3}p_{i}}{(2\pi )^{3}2E_{i}}.
\end{equation}

\section{Numerical results and discussion}

\begin{figure}[b]
\centering
\includegraphics[height=5.5cm,width=0.48\textwidth]{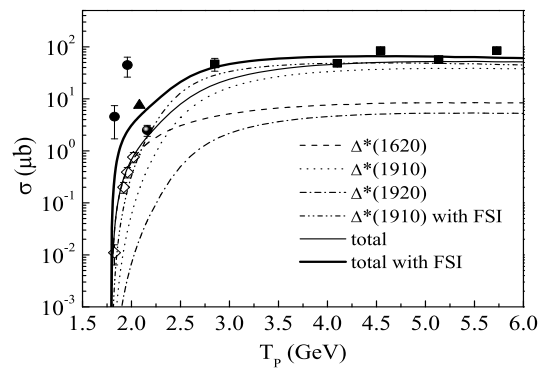}
\caption{The calculated total cross section versus $T_{p}$ for the $%
pp\rightarrow nK^{+}\Sigma ^{+}$ reaction in solution I compared to the data from old
measurement (solid squares) \protect\cite{ab88}, COSY-11 (solid circles)
\protect\cite{tr06}, COSY-ANKE (hollow diamonds and solid ball) \protect\cite{yu07,yu10,yu11},
and HIRES (solid traingles) \protect\cite{ab10}. The dashed,
dotted and dash-dotted curve are contributions from the $\Delta ^{\ast
}(1620)$, $\Delta ^{\ast }(1910)$ and $\Delta ^{\ast }(1920)$, respectively.
The dash-dot-dotted curve is the contribution of $\Delta ^{\ast }(1910)$ with
the $n\Sigma ^{+}$ FSI.
The solid and bold curve are the total contribution without and
with the $n\Sigma ^{+}$ FSI, respectively.}
\end{figure}

\begin{figure*}[t]
\begin{minipage}{1\textwidth}
\includegraphics[height=5.0cm,width=0.42\textwidth]{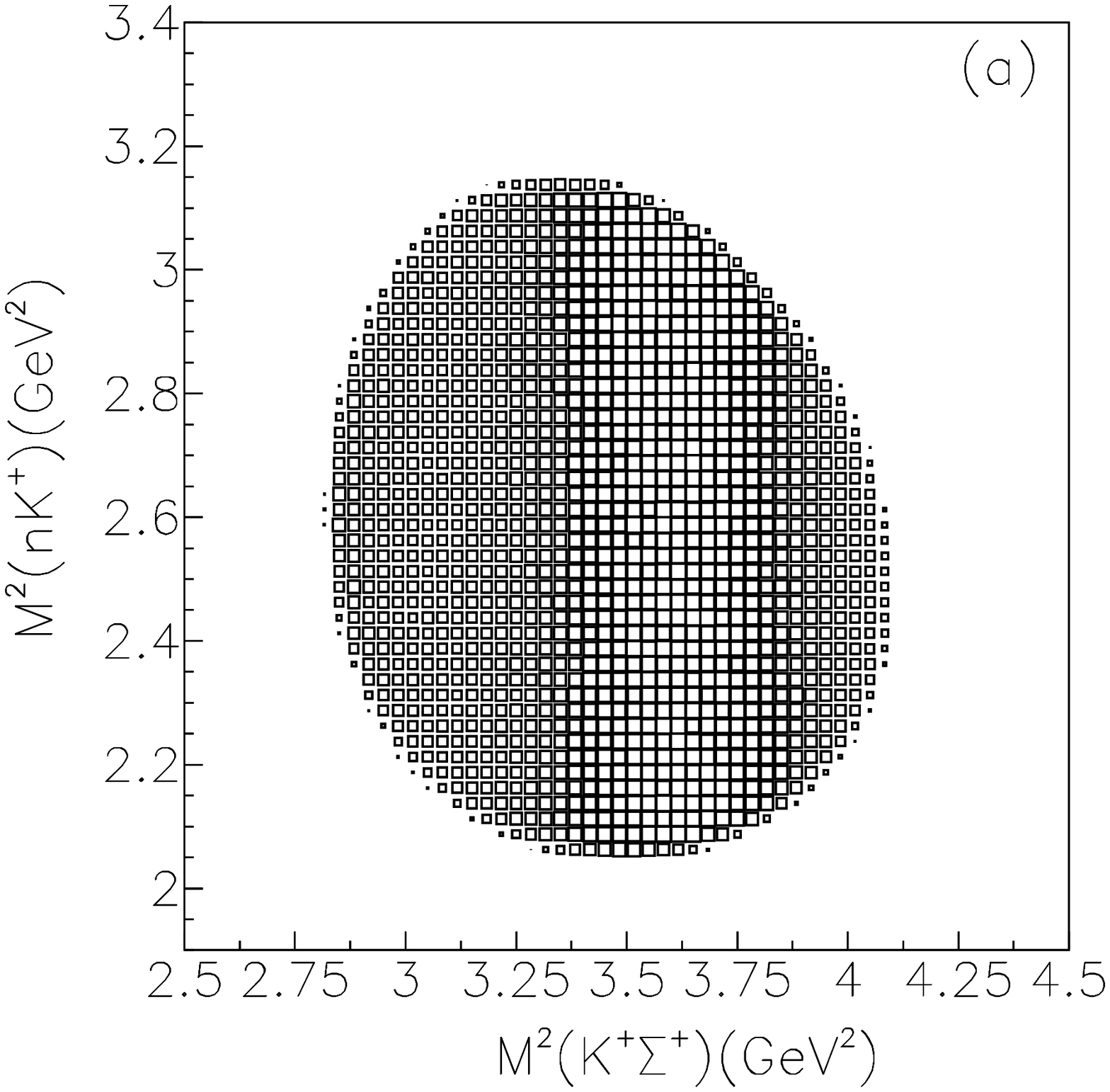}
\includegraphics[height=5.0cm,width=0.42\textwidth]{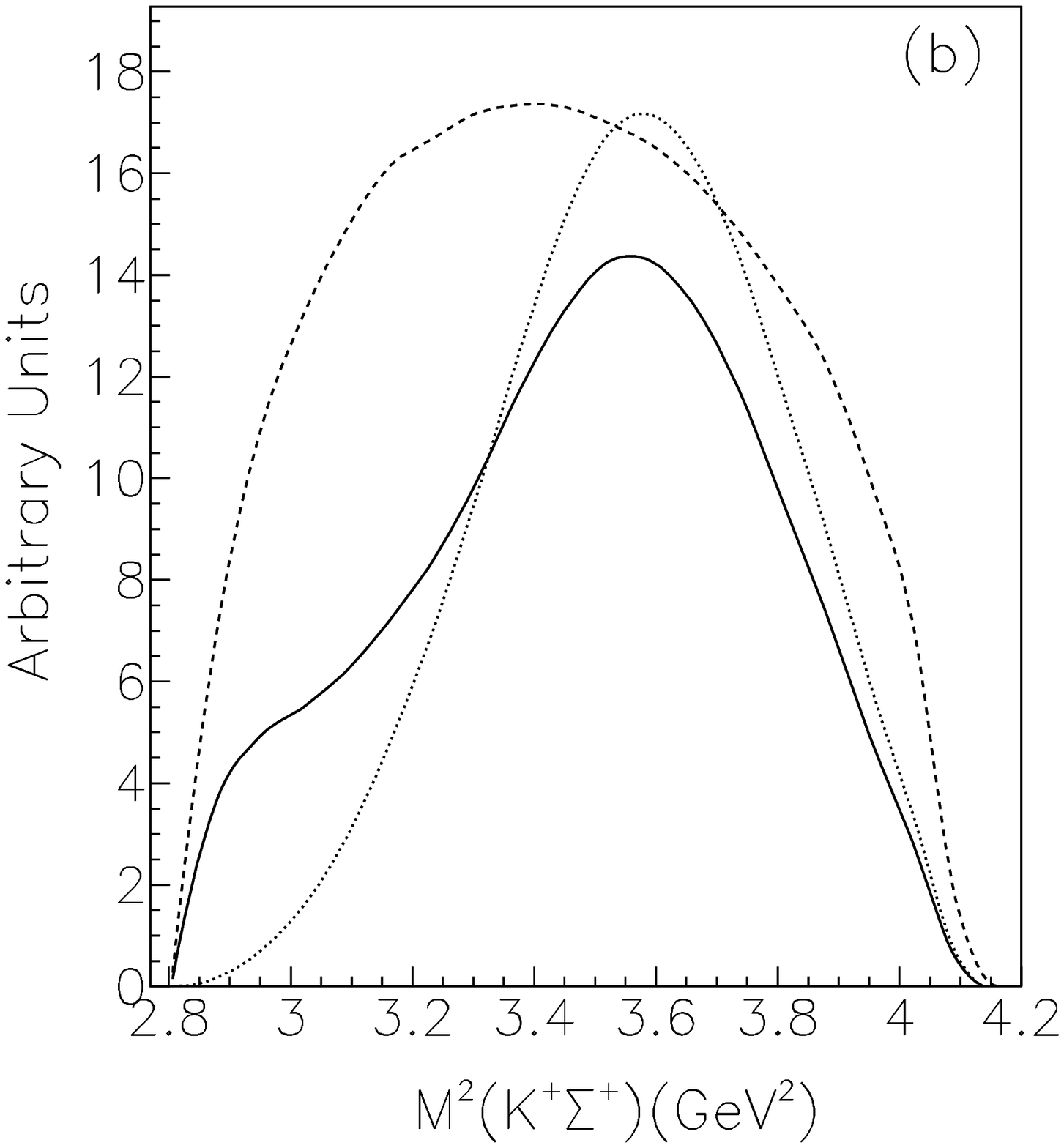}
\end{minipage}
\caption{The Dalitz plot (a) and invariant mass spectrum (b) for the $%
pp\rightarrow nK^{+}\Sigma ^{+}$ reaction at beam energy T$_{p}=$ $2.8$ GeV
in solution I without FSI. The solid line is the total contribution of the $%
\Delta ^{\ast }(1620)$, $\Delta ^{\ast }(1910)$ and $\Delta ^{\ast }(1920)$
resonances and the dotted line is that with $\Delta ^{\ast }(1620)$ turned off.
The dashed curve denote the pure phase space distribution.}
\end{figure*}

With the FOWL code in the CERN program library, the proton beam energy (T$%
_{p}$) dependence of the total cross sections for the $pp\rightarrow
nK^{+}\Sigma ^{+}$ reaction are calculated. As we have mentioned in Sec. II,
we proposed two solutions to interpret the role and contribution of $%
\Delta ^{\ast++ }$ resonances in $pp\rightarrow nK^{+}\Sigma ^{+}$ reaction.
In this section, Fig. 4 $\sim$ 5 present the numerical results of solution I
and Fig. 6 $\sim$ 7 are the calculations for the solution II.

In the solution I as shown in Fig. 4, it is found that the $\Delta ^{\ast
}(1910)$ resonance is dominant at high energy. The contribution of $\Delta ^{\ast }(1920)$
resonance are presented to be negligible, which is consistent with the
results in Ref. \cite{xie07,cao08}.
In the very close-to-threshold energies, the contribution mainly comes from the $\Delta ^{\ast
}(1620)$ resonance. It is noted that the contribution from the $\Delta
^{\ast }(1620)$ is not as large as the calculations in Ref. \cite%
{xie07,cao08} and nearly one order smaller than that of the $\Delta
^{\ast }(1910)$ at the beam energy T$_{p}>2.5$ GeV, because we use smaller
coupling constant of $\Delta ^{\ast }(1620) K\Sigma $ and cut-off
in the form factors. The total contribution from these three
resonances (see the amplitude in Eq. (\ref{eq:solution1})) are in good
agreement with the COSY-ANKE data \cite{ab88,yu07,yu10}. However,
the role of the $\Delta ^{\ast }(1620)$ could be replaced by the strong $n\Sigma^+$ FSI,
see the dash-dot-dotted curve in Fig. 4. If the $\Delta^{\ast }(1620)$ and strong $%
n\Sigma ^{+}$ FSI are both included in the model (see Eq. (\ref{eq:solution1FSI})),
the HIRES data \cite{ab10} could be fitted, as can bee seen by the bold curve in Fig. 4.

At the near threshold region, the Dalitz Plot and invariant mass spectra
are close to the distributions of pure phase space so they give us little information.
The measurements at higher energies can give us more clue of contributing resonances.
Since the kinetic energy of the proton beam T$_{p}$ can reach up to about $%
2.8$ GeV at COSY, we calculate the Dalitz Plot and invariant mass spectra at
T$_{p}=$ $2.8$ GeV. The Fig. 5 depicts our model prediction of the Dalitz plot
and invariant mass spectra in solution I of the amplitude without $%
n\Sigma ^{+}$ FSI in Eq. (\ref{eq:solution1}). In Fig. 5(b), we notice
that there is a bump for invariant mass spectra in the range of $2.8$ GeV $%
^{2}<$ M$^{2}(K^{+}\Sigma ^{+})<3.2$ GeV $^{2},$ which comes from the
contribution of $\Delta ^{\ast }(1620)$ resonance. So if invariant mass spectra
could be the measured with good precision, the role of $\Delta ^{\ast }(1620)$
resonance in the $K\Sigma$ production would be clarified.

\begin{figure}[b]
\centering
\includegraphics[height=5.5cm,width=0.48\textwidth]{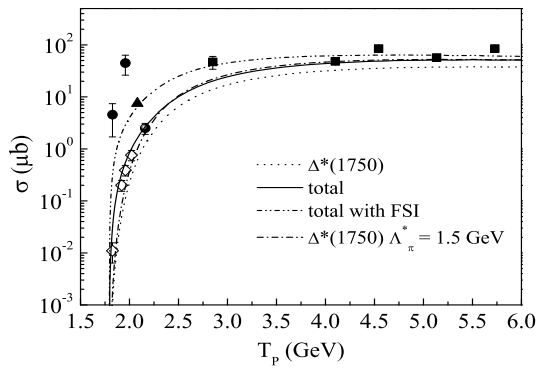}
\caption{The calculated total cross section versus T$_{p}$ for the $%
pp\rightarrow nK^{+}\Sigma ^{+}$ reaction in solution II. The data are the same
as those in Fig. 4. The dotted and dash-dotted
curve are the contribution from the $\Delta ^{\ast }(1750)$ with
$\Lambda _{\pi }^{\ast } = 1.0$ and 1.5, respectively.
The dash-dot-dotted and solid curves are the total
contribution with and without the $n\Sigma ^{+}$ FSI,
respectively.}
\end{figure}

\begin{figure*}[tbph]
\begin{minipage}{1\textwidth}
\includegraphics[height=5.0cm,width=0.42\textwidth]{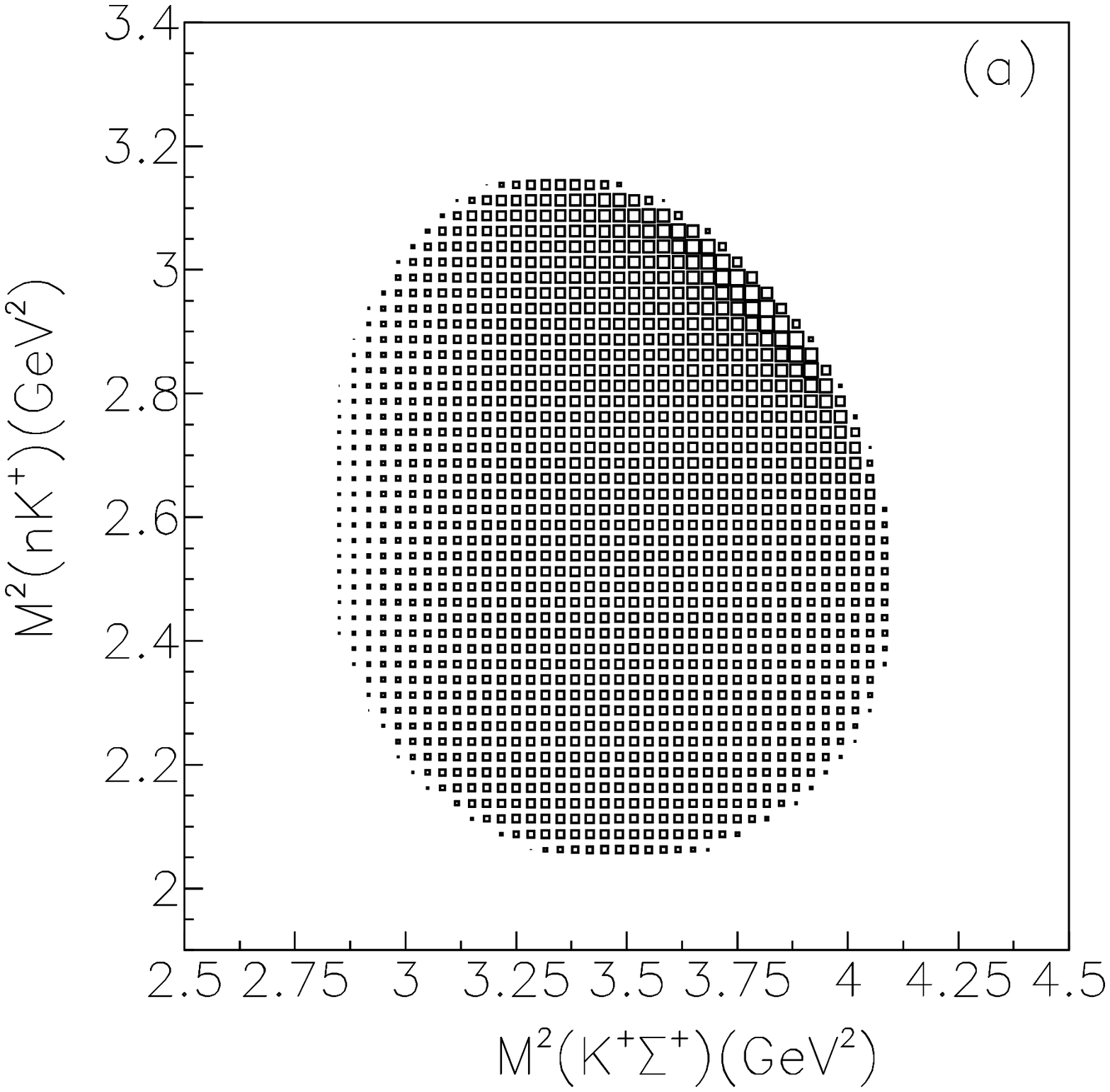}
\includegraphics[height=5.0cm,width=0.42\textwidth]{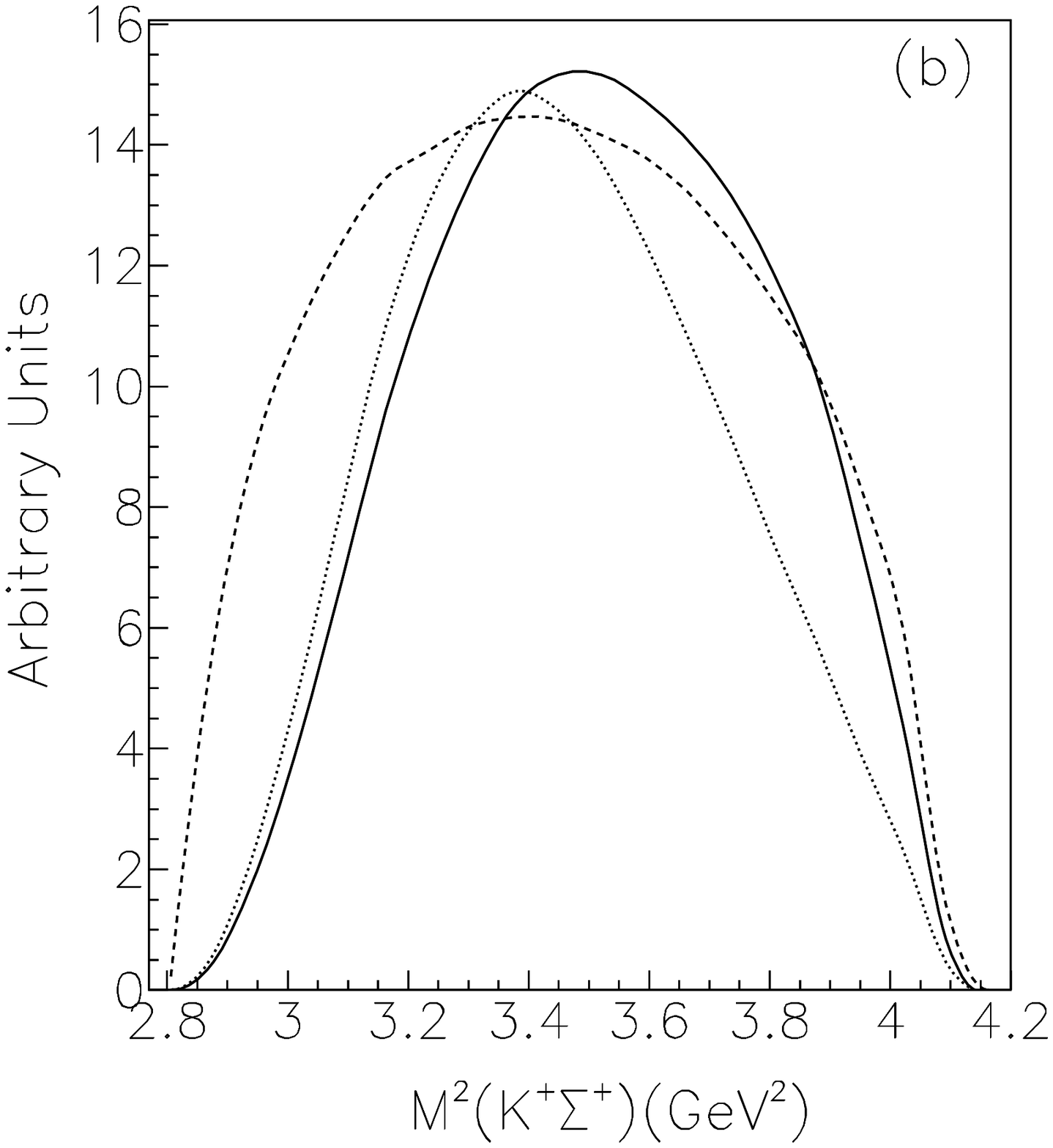}
\end{minipage}
\caption{The Dalitz plot (a) and the solid curve in invariant mass spectrum (b) for the $%
pp\rightarrow nK^{+}\Sigma ^{+}$ reaction at beam energy $T_{P}=$ $2.8$ GeV
with the contributions from the $\Delta ^{\ast }(1750)$ resonance with FSI.
The dotted curve is the $\Delta ^{\ast }(1750)$ resonance without FSI, and
the dashed curve denotes the pure phase space distributions. }
\end{figure*}

In Fig. 6, we present the total cross sections for the $pp\rightarrow
nK^{+}\Sigma ^{+}$ reaction in our solution II.
We find that the calculations with the amplitude in Eq. (\ref{eq:solution2})
can reproduce the COSY-ANKE data \cite{ab88,yu07,yu10} quite well
in the whole energy range. Here we use the same parameters for the
$\Delta ^{\ast }(1620)$ and $\Delta ^{\ast }(1920)$ resonances with those in solution I.
Similar to the solutuon I, the $\Delta ^{\ast
}(1620)$ resonance is important in the very close-to-threshold energies
and the contribution of $\Delta ^{\ast }(1920)$ is small.
The $\Delta ^{\ast}(1750)$ takes the place of the
$\Delta ^{\ast}(1910)$ and dominates at high energies.
As a result, it is seemed that either $\Delta ^{\ast }(1910)$ or $\Delta ^{\ast}(1750)$
can describe the data well and the total cross sections
can not resolve the mystery of mass position of $P_{31}$ resonance.
Meanwhile, the total contribution with strong $n\Sigma ^{+}$ FSI
(see Eq. (\ref{eq:solution2FSI})) describe the HIRES data with good quality \cite{ab10}.
Moreover, it is worthy of attention that the contribution of $\Delta ^{\ast }(1750)$
resonances alone with appropriate $\Lambda _{\pi }^{\ast } = 1.5$ GeV can
describe the COSY-ANKE or HIRES data with or without strong $n\Sigma^+$ FSI,
respectively, as shown in Fig. 6. This reflects the fact that the the role of sub-threshold
$\Delta ^{\ast }(1620)$ resonance is very uncertain cosidering the present
 total cross section data if the dominant $P_{31}$ state is $\Delta ^{\ast }(1750)$.
Fortunately, it would be studied
in the invariant mass spectra, as pointed out above.
Anyway, the HIRES data indicate strong $n\Sigma ^{+}$ FSI in both solutions.

In Fig. 7, we give the Dalitz plot and invariant mass spectra for the $%
pp\rightarrow nK^{+}\Sigma ^{+}$ reaction at T$_{p}=$ $2.8$ GeV with the
contribution of only $\Delta ^{\ast }(1750)$ resonance with $\Lambda _{\pi }^{\ast } = 1.5$.
The influence of the $\Lambda _{\pi }^{\ast }$ on these observables
is minor. Comparing with Fig. 5, we can see that the two schemes, the dominance of $%
\Delta ^{\ast }(1910)$ or $\Delta ^{\ast }(1750)$, are obviously distinguishable.
So we expect the new measurement of the invariant mass spectrum of the $%
pp\rightarrow nK^{+}\Sigma ^{+}$ reaction at high energies could clarify the controversial
spectrum of the $\Delta ^{\ast }$ resonances. Meanwhile, the influence of
the $n\Sigma ^{+}$ FSI is mainly on the invariant mass spectra M$^{2}(n\Sigma ^{+})$ but
the $\Delta ^{\ast }(1620)$ resonance is more obvious in the M$^{2}(K^+\Sigma ^{+})$, so
they can be discriminated in the Dalitz plot and invariant mass spectra as well.

\section{Summary}

The mass of the $P_{31}$ state with isospin 3/2 is highly questionable at
present. Though the $\Delta ^{\ast }(1910)$ resonance is a four-star state
in PDG \cite{pdg} but it is missing in the dynamical coupled-channels
analyses of Excited Baryon Analysis Center (EBAC) at JLab \cite{ns10},
together with another four-star $P_{33}$ state $\Delta ^{\ast }(1920)$.
In their updated analyses which include more channels,
the $\Delta ^{\ast }(1910)$ resonance appear~\cite{Kamano13,Kamano2pi}.
The only $P_{31}$ $\Delta ^*$ state in Giessen model \cite%
{gp02,Penner2,vs04,cao03} is the $\Delta ^{\ast }(1750)$, and it is also
seen in the old KSU analysis \cite{dm92} and Pitt-ANL model \cite{tp00}. The
GWU analysis find one $P_{31}$ pole at $M=1771$ MeV but assigned it as the $%
\Delta ^{\ast }(1910)$ resonance due to its Breit-Wigner mass located at
above 2.0 GeV \cite{ra06}. The Juelich model find a dynamical generated $%
P_{31}$ state around 1750 MeV besides the genuine $\Delta ^{\ast }(1910)$
resonance \cite{am00,am01,md11,Doring12}. However, the $\Delta ^{\ast
}(1750) $ is only a one-star state in PDG \cite{pdg}. The above situation
show that we still have not enough knowledge of these $\Delta ^{\ast }$
resonances. Our calculations in this paper would be helpful for understanding
them better.

In this work, we have calculated the contributions from the $\Delta ^{\ast
}(1620)$, $\Delta ^{\ast }(1750)$, $\Delta ^{\ast }(1910)$ and $\Delta
^{\ast }(1920)$ in the $pp\rightarrow nK^{+}\Sigma ^{+}$ reaction and given
two solutions to interpret the role and contribution of the $\Delta ^{\ast
++}$ resonances in this reaction based on the present data of total cross sections.
In solution I, the contribution from the $P_{31}$ $\Delta ^{\ast }(1910)$ resonance
is dominant at high energies. In solution II, we find another $P_{31}$ state
$\Delta ^{\ast }(1750)$ above threshold is most important,
by combining with the experimental data of $\pi
^{+}p\rightarrow K^{+}\Sigma ^{+}$ reaction. The present close-to-threshold data of total cross
sections can not pin down that the $P_{31}$ state is $%
\Delta ^{\ast }(1750)$ or $\Delta ^{\ast }(1910)$.
Only after the mass of the main resonance is determined,
the remaining free parameters, namely the decay ratios of resonances and
cut-off in the form factors, will be well determined by the measured data.
Then the mechanism of $K\Sigma$ production would be explained with more confidence.
At present, it is difficult to give a detailed error analysis of our model.

More seriously, the inconsistent close-to-threshold data from several groups
result in the rather inconclusive status of the contribution at low
energies. Either the sub-threshold $P_{31}$ $\Delta ^{\ast }(1620)$ resonance or
strong $n\Sigma^+$ FSI or both is possibly significant at close to threshold
region. If the HIRES data is only an upper bound of the total cross section
as argued by Valdau and Wilkin \cite{yu11}, we can conclude that the $\Delta
^{\ast }(1620)$ would be strongly coupled to the $K\Sigma$ if the $\Delta
^{\ast }(1910)$ is responsible for the $K\Sigma$ production at high
energies. However, if the strong coupling of the $\Delta ^{\ast }(1750)$ to
the $K\Sigma$ is confirmed, it is probable that the strong $n\Sigma^+$
interaction is excluded to some confidential level and the coupling of the $%
\Delta ^{\ast }(1620)$ to the $K\Sigma$ has to be checked by
the low range of M$^{2}(K^+\Sigma ^{+})$ in invariant mass spectra.

Fortunately, it is hopeful that the invariant mass distributions and the
Dalitz Plot could discriminate these solutions because various contribution
is evidently distinguishable as we have presented. Though the experiment
would be challenging because of the neutron in the final states, it is
encouraging to measure these observables in the future considering the very
controversial location of the $\Delta ^*$ resonance and their coupling to
the $K\Sigma$ channel.

\section{Acknowledgments}

The author X. Y. Wang is grateful for Dr. Qing-Yong Lin for valuable
discussions and help. This project is partly supported by the National Basic
Research Program (973 Program Grant No. 2014CB845406) and the National
Natural Science Foundation of China (Grant Nos. 11347156, 11405222, 11105126 and 11475227). We
acknowledge the one Hundred Person Project of Chinese Academy of Science
(Y101020BR0).

\end{document}